# Cometary Plasma Science

A White Paper in response to the Voyage 2050 Call by the European Space Agency

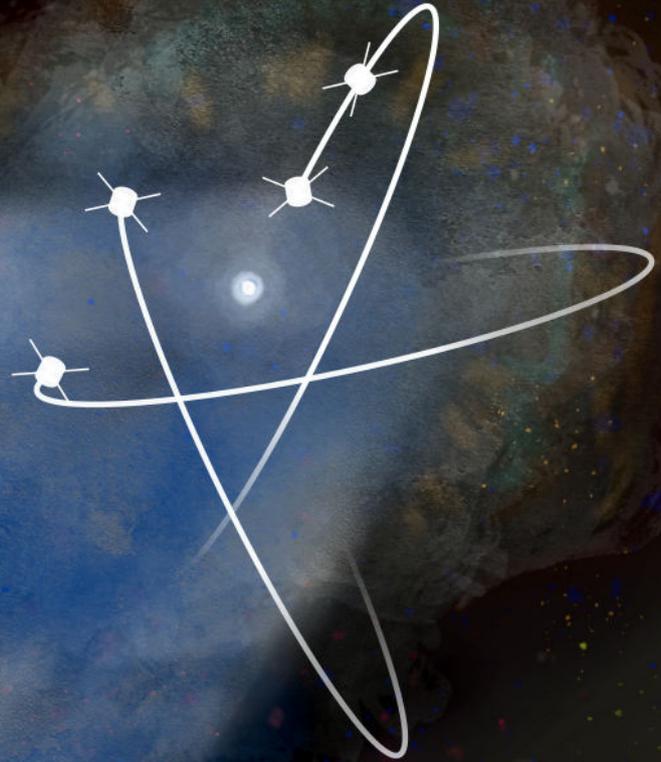

Charlotte Götz

c.goetz@tu-bs.de

TU Braunschweig
Mendelssohnstr. 3
38106 Braunschweig
Germany

# Abstract

Comets hold the key to the understanding of our solar system, its formation and its evolution, and to the fundamental plasma processes at work both in it and beyond it. A comet nucleus emits gas as it is heated by the sunlight. The gas forms the coma, where it is ionised, becomes a plasma and eventually interacts with the solar wind. Besides these neutral and ionised gases, the coma also contains dust grains, released from the comet nucleus.

As a cometary atmosphere develops when the comet travels through the solar system, large-scale structures, such as the plasma boundaries, develop and disappear, while at planets such large-scale structures are only accessible in their fully grown, quasi-steady state. In situ measurements at comets enable us to learn both how such large-scale structures are formed or reformed and how small-scale processes in the plasma affect the formation and properties of these large scale structures. Furthermore, a comet goes through a wide range of parameter regimes during its life cycle, where either collisional processes, involving neutrals and charged particles, or collisionless processes are at play, and might even compete in complicated transitional regimes. Thus a comet presents a unique opportunity to study this parameter space, from an asteroid-like to a Mars- and Venus-like interaction.

The Rosetta mission and previous fast flybys of comets have together made many new discoveries, but the most important breakthroughs in the understanding of cometary plasmas are yet to come. The Comet Interceptor mission will provide a sample of multi-point measurements at a comet, setting the stage for a multi-spacecraft mission to accompany a comet on its journey through the solar system.

This white paper reviews the present-day knowledge of cometary plasmas, discusses the many questions that remain unanswered, and outlines a multi-spacecraft ESA mission to accompany a comet that will answer these questions by combining both multi-spacecraft observations and a rendezvous mission, and at the same time advance our understanding of fundamental plasma physics and its role in planetary systems.

# Contents

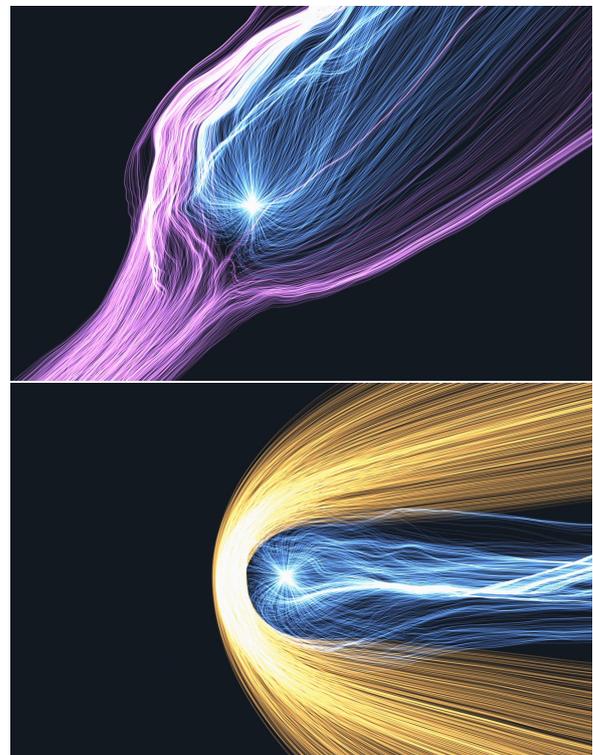

*Simulations of comet plasmas. Top: stream lines of solar wind and cometary ions. Bottom: magnetic field lines and cometary ion stream lines. (Technische Universität Braunschweig and Deutsches Zentrum für Luft- und Raumfahrt; Visualisation: Zuse-Institut Berlin)*

# 1  Present day knowledge

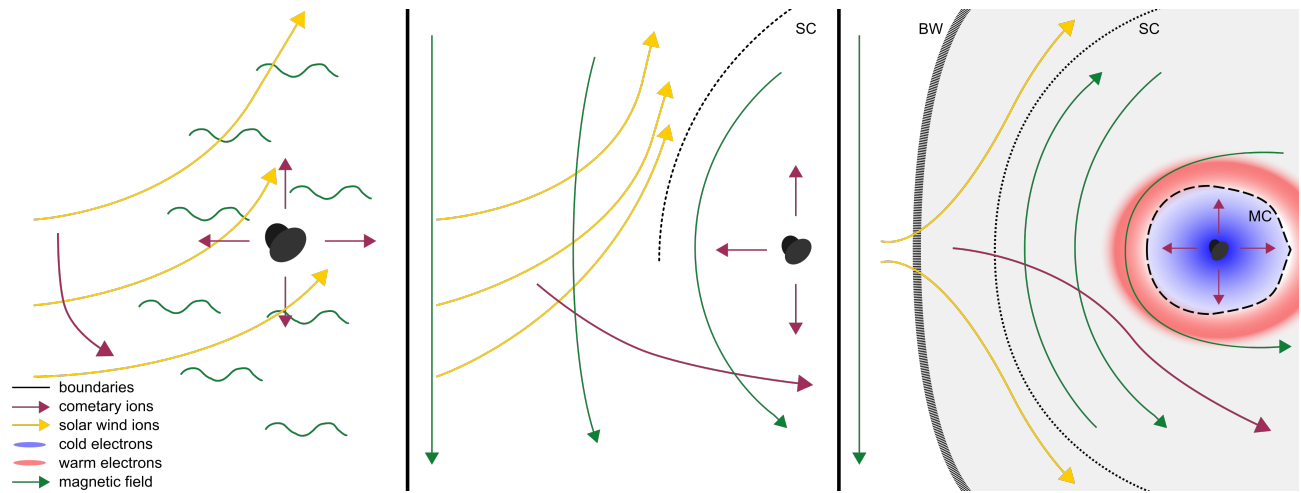

Figure 1.1: *A sketch of the cometary plasma environment in the plane containing the magnetic field and the solar wind flow. The three panels show different stages, left: weak activity, middle: intermediate activity, right: high activity. Boundaries and regions are labelled: bow wave (BW), solar wind ion cavity (SC), and diamagnetic cavity (MC). Adapted from Götz (2019).*

At a comet there are two plasma types: (1) the light solar wind ions, and (2) the cometary plasma, which usually consists of heavy ions that are produced from the ionisation of the neutral gas that surrounds a comet nucleus. As the neutral gas (called the *coma*) is not gravitationally bound due to the small size of the nucleus (100 m to 100 km), the neutrals and ions have a small radial velocity. Unhindered they would expand indefinitely into the near-vacuum of space. However, Biermann (1951) found that the interaction of the cometary ions with the solar wind could accelerate them and form the plasma tail structures that are observable, sometimes even by eye, from Earth.

Comets can behave similarly to Mars and Venus in their **interaction with the solar wind**, since at both those planets the main obstacle is the conductive atmosphere and not the planets themselves nor their magnetic fields. This is also the case at comets, although there are significant differences in the outgassing speeds due to the much larger gravity of terrestrial planets and obstacle size difference.

The interaction of the two different types of plasma, protons from the solar wind and usually water or carbon dioxide ions from the comet, has been an object of studies for many years now. The spacecraft encounters with comets 21P/Giacobini-Zinner (21P) and 1P/Halley (1P) in the 1980s heralded the advent of modern cometary plasma science (Riedler et al., 1986). Beforehand, only remote observations were available. The golden age of cometary plasma science began with the arrival of the Rosetta spacecraft at comet 67P/Churyumov-Gerasimenko (67P, Glassmeier et al., 2007), which was the first spacecraft to orbit a comet and take detailed measurements of the environment for an entire perihelion passage.

In principle, the interaction of the two flows (cometary and solar wind) can be described by the idea of a **mass-loaded plasma**. The addition of the slow, heavy cometary ions to the high velocity solar wind leads to the modification of both the cometary and solar wind plasmas. The degree to which the solar wind is mass-loaded depends on the number of cometary ions produced per second, which in turn depends on the outgassing rate of the cometary neutrals. Both are mostly anti-correlated with the heliocentric distance of the comet, although there are also be small-scale variations depending on other parameters.

A magnetohydrodynamic approach (multi- or single-fluid) for the mass-loaded plasma can approximate the interaction region, but leaves many effects at scales smaller than the ion gyroradius unsolved. At comets, the ion gyroradius is often much larger than typical length scales. In some instances, even electron scales need to be taken into account to understand large-scale features (Deca et al., 2017).

Ordered by the neutral outgassing rate $Q$, the interaction generally falls into one of three regimes, which are illustrated in Figure 1.1.

**The strongly active comet:** $Q > 5 \times 10^{27} \text{s}^{-1}$   This is the classical comet plasma picture as it was known from the missions to comet 1P, 19P/Borelly, 21P and 26P/Grigg-Skjellerup (26P). Boundaries like the bow shock and diamagnetic cavity have formed (Neubauer et al., 1986). The inner coma is closer to photo-chemical equilibrium and collisions between ions and neutrals are important. There is a visible plasma tail. There are waves far upstream of the bow shock from the pickup of cometary ions (Coates and Jones, 2009).

**The intermediately active comet:** $5 \times 10^{26} \text{s}^{-1} < Q < 5 \times 10^{27} \text{s}^{-1}$   At this stage, the solar wind is deflected and decelerated significantly, as a result of the presence of cometary ions. First boundaries form, but can disappear and reform on short timescales (Gunell et al., 2018a). The interplanetary magnetic field starts to drape around the obstacle ion cloud.

**The weakly active comet:** $Q < 5 \times 10^{26} \text{s}^{-1}$   No boundaries have formed yet. The influence of cometary ions on the solar wind is small. The magnetic field is usually only slightly elevated compared to solar wind values (Goetz et al., 2017) and the plasma density follows a typical $1/r$ profile that is modulated by the neutral outgassing rate (Edberg et al., 2015; Galand et al., 2016). There are ultra-low frequency waves detected in the magnetic field and plasma density (Breuillard et al., 2019; Richter et al., 2015).

A comet, along its journey around the Sun, may move to higher outgassing regimes while others remain weakly active throughout. Comet 67P went through all three stages listed above during the Rosetta mission (Hansen et al., 2016).

## 1.1  A review of large scale structures in the interaction region

### 1.1.1  Bow Shock

**The bow shock has been observed at several active comets** and has been modelled extensively (Koenders et al., 2013). Figure 1.1 shows schematic images of solar wind interaction with a comet, and the bow shock appears in the right-hand panel. The bow shock moves outwards as the mass-loading increases, and can be millions of km from the nucleus in comets with high gas production rates, whereas at low gas production rates (e.g. 26P) the critical point for shock formation is never reached and no bow shock forms; instead a more gradual increase in magnetic field (a bow wave) is observed (Scarf et al., 1986). At comet Halley, the bow shock was observed by Giotto only on the inbound pass, and a bow wave could be observed outbound. At comet 67P, the trajectory of the Rosetta spacecraft did not allow for an in-situ observation of a bow shock or bow wave, but a structure in the plasma environment at lower gas production rates was identified as an infant bow shock, a highly asymmetric structure that behaves like a shock and is confined to one hemisphere of the interaction region, as illustrated in the middle panel of Figure 1.1 (Gunell et al., 2018a).

**Bow shocks are not unique to the comet plasma environment**, and they have also been seen at all planets. At Mars the bow shock is largely symmetric, and its mean location is steady and only weakly affected by solar cycle variations (Mazelle et al., 2004). For both Mars and Venus the position of the bow shock has been found to be more influenced by solar Extreme UltraViolet (EUV) radiation than by solar wind dynamic pressure (Hall et al., 2016; Shan et al., 2015).

The bow shock at a comet reacts to increased ionisation rates in the same way as the bow shocks at Mars and Venus. It has been shown in simulations that the standoff distance of a cometary bow shock increases with an increasing ionisation rate. The more realistic simulations are made by including additional ionisation processes — photo-ionisation, electron–impact ionisation, and charge exchange — the farther upstream from the nucleus the bow shock moves (Simon Wedlund et al., 2017). The acceleration of newly created pickup ions differs on the upstream and downstream sides of the shock. Therefore a pickup ion energy spectrum can be used to estimate the standoff distance of a bow shock, as was shown in simulations (Alho et al., 2019) and this was used to estimate the position of the bow shock at comet 67P when the spacecraft was located far downstream (Nilsson et al., 2018).

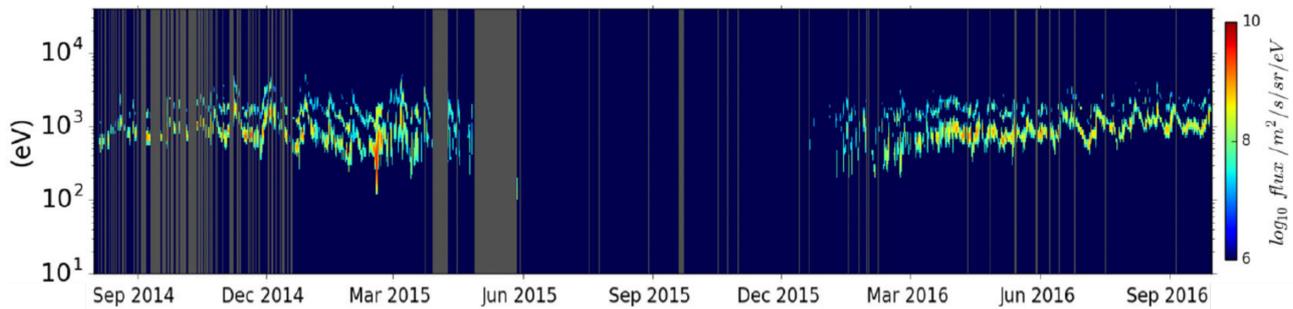

**Figure 1.2:** *Energy spectrogram of the solar wind ions summed over all viewing directions and integrated over 1 hour for the entire Rosetta comet phase. One can see clearly that the spacecraft was located in a solar wind ion free region during the months around comet 67P perihelion (August 2015), when the gas production rate was highest. Adapted from Nilsson et al. (2017).*

A multitude of upstream wave phenomena have been observed at both Mars, Earth, and Venus (e.g. Delva et al., 2015; Kempf et al., 2015; Mazelle et al., 2004). Waves were also observed in the foreshock of comet 1P (Oya et al., 1986), and future detailed observations in the upstream region of a cometary bow shock would be expected to show similar features. This would include back-scattered particles that contribute to wave growth, which has been seen at all three of the terrestrial planets.

### 1.1.2 Solar Wind Ion Cavity

When comet 67P was about 1.8 AU from the Sun, solar wind ions could no longer reach the inner coma (Nilsson et al., 2017). The **region that is devoid of solar wind ions** is called the solar wind ion cavity (Behar et al., 2017; Simon Wedlund et al., 2019a). Closely upstream of the solar wind ion cavity, **solar wind ions are seen to be significantly deflected** from their original anti-sunward motion, and protons back-scattered toward the sun have also been detected (Behar et al., 2017). The location of this region changes with gas production rate and upstream solar wind parameters. For example, it was observed at comet 67P, that an interplanetary coronal mass ejection (ICME) could push solar wind ions closer to the inner coma, so that Rosetta, previously in the solar wind ion cavity, could observe protons for a short period of time (Edberg et al., 2016). A region from which the solar wind was excluded was also seen at comet 26P (Johnstone et al., 1993), and at comet Halley. The boundary that separates the solar wind ions from the solar wind ion cavity at comet 1P has been given many different names in the literature depending on what aspect of it was under study. For example the term *cometopause* has been used both to describe where cometary ions dominate over solar wind ions (Galeev et al., 1988) and to mean a solar wind charge-exchange collisionopause (Cravens, 1989). A similar boundary has been called the Induced Magnetosphere Boundary at Mars (Lundin et al., 2004) and Venus (Zhang et al., 2008). More than one physical mechanism is likely to be involved in its formation, as both collisional effects, magnetic pileup and ion chemistry are important in this region. See Coates and Jones (2009) for a review of the many aspects relevant to this boundary at comets.

### 1.1.3 Diamagnetic Cavity

Early on in cometary plasma physics, Biermann et al. (1967) and Galeev et al. (1985) realised that one consequence of mass-loading is the deceleration of the incoming (solar wind) flow. The ultimate consequence of this is that the flow comes to a halt at some cometocentric distance $r_c$ if mass-loading is sufficient. Although the magnetic field was not part of their simple fluid models, they realised that as long as the magnetic field was frozen into the flow, it would also stop at this distance. As comet nuclei are not magnetised (Auster et al., 2015), it was speculated that a field-free diamagnetic cavity would form. However, this region can only be sustained if the magnetic field diffusion into it is prevented. Early on it was speculated that simply the dynamic pressure of the outflowing cometary ions would be sufficient to balance the magnetic pressure and prevent diffusion into the cavity.

In preparation for the space missions to comet 1P/Halley, an **artificial comet experiment** (AMPTE, Valenzuela et al., 1986) was designed and launched. The main goal was to investigate the interaction of the solar wind and magnetospheric plasma with a cloud of heavy ions, in this case Barium or Lithium. The plasma parameters of these experiments were very similar to the parameters during the 1P flybys.

In regards to the formation mechanism of the diamagnetic cavity at the artificial comet, two models have been presented: Haerendel et al. (1986) showed that the dynamic pressure of the expanding ion cloud is sufficient to stave off the magnetic field, whereas Valenzuela et al. (1986) and Luehr et al. (1988) showed that the thermal pressure of the electrons could also be responsible. Sauer and Baumgaertel (1987) showed that in numerical simulations, the dynamic pressure was the more favourable of the two mechanisms. No other studies were conducted and so far, none of the two mechanisms could be ruled out entirely.

With the 1P flyby of the European Space Agency's Giotto spacecraft, new light was shed on the **diamagnetic cavity shape and formation mechanism**. It was quickly found that neither the thermal pressure nor the dynamic pressure would be sufficient to uphold the diamagnetic cavity, because neither of them showed a significant change at the boundary (Cravens, 1986). So

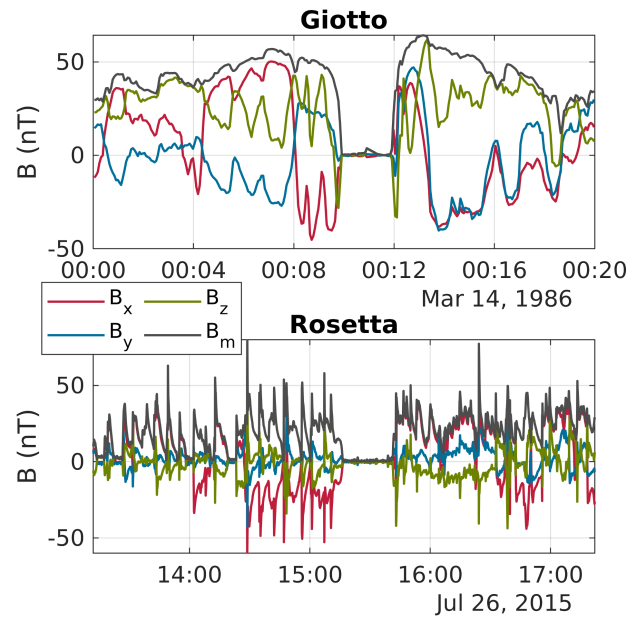

Figure 1.3: *Magnetic field observations of cavities at 1P (measured by Giotto's magnetometer) and 67P (measured by the magnetometer onboard Rosetta).*

Cravens (1986, 1987) presented an alternative mechanism: the **ion–neutral friction force**. For this mechanism it was assumed that the magnetic field in the upstream region had already reached a stagnation point and the charge-exchange collisions between the outward streaming neutrals and the ions at rest could balance the magnetic field pressure. This of course assumes that the ion–neutral coupling is efficient, which was consistent with observations of the ion and neutral speeds being very similar. Cravens (1987) derived a magnetic field profile in the boundary region and a standoff distance for the diamagnetic cavity that fit well with the observations by the Giotto magnetometer.

Neubauer (1987) then pointed out that the diamagnetic cavity boundary was very likely not spherical in shape, as the boundary normal was a better fit to an unstable boundary. Indeed, Ershkovich and Mendis (1986) and Ershkovich and Flammer (1988) found that the boundary might be unstable to the Flute and Kelvin-Helmholtz instabilities. This was later confirmed in simulations by Rubin et al. (2012), who found that the boundary was rippled. However, no measurements had yet confirmed this.

With the arrival of the Rosetta mission at comet 67P, the diamagnetic cavity could be investigated in more detail (Goetz et al., 2016b). The Rosetta spacecraft entered the cavity over 700 times, although it should be noted that because of the negligible speed of the spacecraft this means that the boundary was moving over the spacecraft and not the other way around as was the case of Giotto at 1P. Figure 1.3 shows example magnetic field measurements at both comets.

Goetz et al. (2016a) and Götz (2019) reported that the diamagnetic cavity size was strongly correlated to the local outgassing rate (derived from in-situ measurements of the neutral gas density). As expected, the diamagnetic cavity expands with increasing outgassing rate. It was also found that the boundary normal was inconsistent with a spherical shape, indicating again that the boundary was rippled and highly unstable.

However, it was found that **the ion–neutral friction force was not the driving parameter** behind the cavity formation at comet 67P, as the measurements of the ion velocity indicated that the coupling of ions and neutrals was inefficient due to lower neutral densities at 67P (Odelstad et al., 2018; Vigren et al., 2017). Additionally, the ion density profile that was assumed for comet 1P was not applicable at 67P, due to transport of the ions being as important as recombination (Beth et al., 2018). However, Henri et al. (2017) found that the electron exobase was a good ordering parameter for the diamagnetic cavity detections, indicating the importance of electron–neutral interactions in this regime. As of now the formation mechanism of the diamagnetic cavity at 67P is still unknown.

### 1.1.4 Plasma Tail

Comets can have more than one tail. In addition to the most clearly visible dust tail there is an ion or plasma tail. While the dust grains in the dust tail are pushed away from the Sun by the photon pressure, the sunlight cannot explain the formation of a plasma tail. This led to Biermann (1951) to propose the **existence of a solar wind**, and Alfvén (1957) to develop a theory for how the solar wind magnetic field lines are draped around the comet. Alfvén's theory was supported by observations at comet 21P by Slavin et al. (1986), who observed that "The structure of the 21P magnetotail was quite similar in many respects to that observed at Venus." Plasma tails can have an enormous length of over 3 AU, as evidenced by some of Ulysses' fortuitous comet tail crossings (Gloeckler et al., 2000; Jones et al., 2000). Tangential discontinuities in the solar wind approaching a comet, as seen in the magnetic field measurements in the coma of comet 1P/Halley (Riedler et al., 1986), can lead to a more complicated magnetic structure characterised by "nested draping" in the plasma tail.

Rays of light pointing away from the nucleus over a range of directions are often seen in telescope images of comets (Rahe, 1968) and it has been suggested that the formation of such cometary rays is related to ionisation processes in the coma (Rahe and Donn, 1969). To date there are no in situ observations of cometary rays.

Remote observations sometimes show that the tail pattern is disrupted and **the tail appears broken or disconnected**. Usually a new tail forms quite quickly (Vourlidas et al., 2007). Three different categories of triggers have been proposed: a shock wave, a magnetic field reversal and a high solar wind dynamic pressure event.

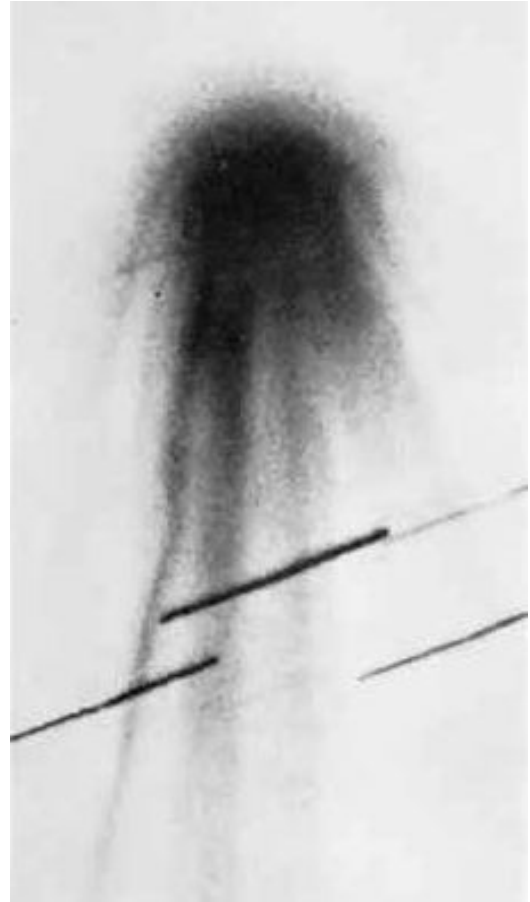

Figure 1.4: *Photograph of Comet Morehouse, (Rahe and Donn, 1969).*

**Shock wave:** Wegmann (1995) proposed that a shock wave travelling down the tail would rarefy and compress the plasma in the tail, which appears to a remote observer as a succession of regions with and without cometary plasma. Thus, in this model, a tail disconnection event is not a real disconnection, it just appears as one to a remote observer.

**Magnetic field reversal:** Niedner et al. (1981) proposed that a field reversal at a discontinuity in the solar wind, like an interplanetary coronal mass ejection, could also trigger reconnection when the discontinuity hits the coma and that this could lead to a disconnection event. Although the reconnection region would be on the dayside, the disturbed plasma may travel towards the tail and cause a real disconnection of the field lines from the inner coma. This would only be possible at very active comets, where the pile-up of the magnetic field is sufficient.

**High solar wind dynamic pressure:** Ip and Mendis (1978) proposed that a flute instability that is triggered in the cometosphere due to higher solar wind dynamic pressure could propagate into the tail and develop into an apparent tail disconnection.

Because tail disconnections occur in the far tail, in-situ observations are difficult to do. Although some far tails have been crossed (e.g. Neugebauer et al., 2007) these observations have been too short to investigate disconnection events. Thus, there have not been any in-situ observations of a tail disconnection, and the cause of tail disconnections remains an open question.

## 1.2 A review of plasma processes at a comet

### 1.2.1 Collisions in the coma

**Solar wind ion interaction with the neutral coma**
In a charge exchange reaction, one or several electrons are semi-resonantly transferred between a neutral particle (atom or molecule) and an ion. Such ion-neutral reactions between incoming, usually fast, ions and a correspondingly slow-moving neutral environment are ubiquitously present in astrophysics environments (Dennerl, 2010; Wargelin et al., 2008). Renewed interest in these reactions was kick-started by the discovery that comets are soft X-ray emitters (Lisse et al., 1996), due to highly-charged solar wind ions charge-exchanging with the neutral coma (Cravens, 1997).

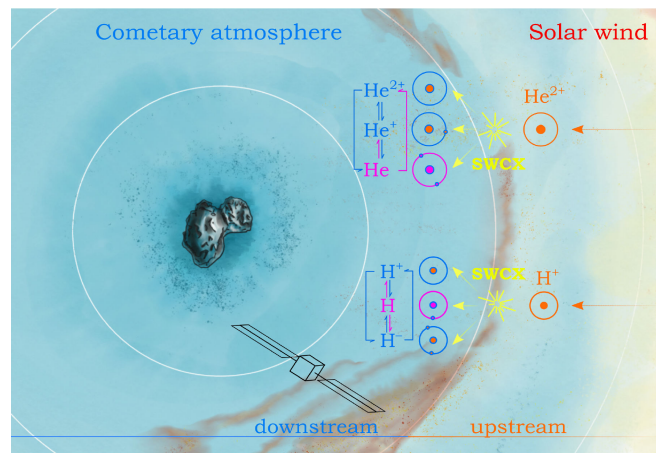

Figure 1.5: *Solar wind charge-exchange interactions at comet 67P (Simon Wedlund et al., 2019b).*

At a comet, single charge-exchange reactions between, for instance, solar wind protons and the neutral gas M take the form $H^+ + M \rightarrow H + M^+$: from the point of view of the ions, the net effect is to **replace a fast, light ion (solar wind ion) with a slow heavy one (newly-born cometary ion)**. Energetic neutral atoms of hydrogen can be a by-product of the reaction (see, e.g. Nilsson et al., 2015). Because the cometary neutral coma is in radial expansion from the nucleus, charge-exchange reactions act cumulatively over distances of hundreds of thousands of kilometres upstream of the nucleus, hence critically contributing to the mass-loading of the plasma, its large-scale dynamics, and to the formation of typical structures such as the bow shock (Gombosi, 1987; Simon Wedlund et al., 2017; Simon Wedlund et al., 2019c).

Slowing-down of solar wind ions due to mass-loading and heating around the shock-like structure ahead of the nucleus are expected, which may call for the use of **energy-dependent cross sections** depending on the severity of these effects (Simon Wedlund et al., 2019a,b,c). The gas production rate of a comet can be estimated, using a model of charge exchange in the coma and in-situ flux measurements of the charge state distribution of solar wind ions (Simon Wedlund et al., 2016; Simon Wedlund et al., 2019a,b,c). This may even lead to measurable X-ray emissions (Häberli et al., 1997).

**Cometary plasma interaction with the neutral coma**   Photo-electrons, produced by solar EUV radiation ionising the neutral coma, are born with energies typically around 10 eV. At comet 67P, a supra-thermal electron population was found, peaking in the 30–40 eV energy range (Broiles et al., 2016). These electrons, in turn, produce secondary electrons below ~12 eV. This picture has been confirmed by plasma measurements, which showed that the bulk of the electron population at comet 67P was warm (5–10 eV) at heliocentric distances above 3 AU (Engelhardt et al., 2018; Eriksson et al., 2017; Gilet et al., 2017).

When comet 67P was at heliocentric distances between 3 AU and perihelion (1.24 AU), there was, in addition to the warm component, a **cold electron population** with temperatures below 1 eV (Engelhardt et al., 2018). This cold population is a result of electron cooling through collisions with the neutral gas (Eriksson et al., 2017; Odelstad et al., 2018). For very high outgassing conditions, for example comet 1P at 0.9 AU, cold electrons have been predicted to be dominant in the inner coma (Gan and Cravens, 1990).

At comet 67P it was found that, while electron impact ionisation dominates plasma production at large heliocentric distances and during transient solar wind events (Hajra et al., 2018), photo-ionisation is the main source of the plasma near perihelion (Heritier et al., 2018b). There, the coma starts to be optically thick to solar EUV radiation due to absorption by the neutral coma (Beth et al., 2018) and the dust (Johansson et al., 2017). Transport was found to be the dominant loss process at comet 67P throughout the Rosetta mission, and dissociative recombination could be significant only close to perihelion (Beth et al., 2018).

Modelling the cometary plasma density, taking both sources and losses into account, showed excellent agreement with Rosetta multi-instrument observations (Galand et al., 2016; Heritier et al., 2018b) at large heliocentric distances (>3 AU), all the way down to the surface (Heritier et al., 2017b). In these models, the ions were assumed to move at the same speed as the neutrals. Therefore, the agreement with data implies that no significant ion acceleration took place within about 70 km from the nucleus. For comet 67P close to perihelion, such models overestimate the plasma density, which indicates that a significant ion acceleration took place in agreement with observations by Odelstad et al. (2018). A presence of nanograins may also have influenced the electron density (Johansson et al., 2017).

For very high outgassing conditions, for example comet 1P at 0.9 AU, the plasma can be assumed to be in photo-chemical equilibrium and ion–electron dissociative recombination would be the dominating loss process (Cravens, 1987).

Ion–neutral collisions are significant in determining **the composition of the plasma**. Charge-changing collisions, which may transfer an electron (e.g., $H_2O^+ + CO_2 \longrightarrow H_2O + CO_2^+$) or a proton (e.g., $H_2O^+ + H_2O \longrightarrow H_3O^+ + HO$) between ions and neutrals, have an influence on both the mass and velocity distributions of the ions. They therefore play a role in mass loading (Szegö et al., 2000). Several ion species were found for low outgassing conditions at comet 67P (Fuselier et al., 2015). While some of these, like $H_2O^+$ and $O^+$, can be produced directly through ionisation, others, like $H_3O^+$, only result from ion–neutral chemistry. Their presence shows that the coma was not fully collisionless. Near perihelion the $H_3O^+$ to $H_2O^+$ ratio was found to be highly variable (Fuselier et al., 2016), and neutral outgassing and ion-neutral collision frequency increased, favouring the production of new ions (Heritier et al., 2017a), in particular those produced by protonation of molecules with higher proton-affinity than that of water (Vigren and Galand, 2013), for instance transforming $H_2O^+$ to $NH_4^+$ in the presence of $NH_3$ (Beth et al., 2016). Changes in the solar wind upstream conditions can change the composition of the neutral coma even on short time scales (Noonan et al., 2018).

### 1.2.2 Electric fields

The three most important contributions to the DC electric field in the inner coma are the solar wind **convectional** electric field, the **ambipolar** field, and the **polarisation** electric field.

In the inner coma, the electrons are hotter than the ions and can escape much faster radially outward from the nucleus. This creates an **ambipolar** electric field (directed radially outward) that accelerates the ions and slows down the outward motion of the electrons. Vigren and Eriksson (2017) have shown that the presence of an electric field can dominate over the effects of collisions and result in much higher ion velocities than predicted based on measurements at 1P. This was confirmed by Langmuir probe measurements in and near the diamagnetic cavity of comet 67P (Odelstad et al., 2018). Also for low outgassing conditions, the ion motion can be faster than the neutral motion as a result of convective and ambipolar fields acting on the ions (Beth and Galand, 2018; Koenders et al., 2016). However, this is only the case for larger cometocentric distances, close to the nucleus (<10 km) no acceleration could be observed (Heritier et al., 2017b).

In the inner coma of a weakly outgassing comet, the ions are unmagnetised and therefore, water group ions, newly created by ionisation, move in the direction of the electric field. Electrons, on the other hand, are magnetised and their motion is governed by an $\vec{E} \times \vec{B}$ drift perpendicular to both the magnetic and electric fields. This leads to a charge separation, which in turn gives rise to a polarisation electric field (Nilsson et al., 2018). Particle-in-cell simulations including all three field contributions have confirmed the existence of a polarisation field in agreement with Rosetta observations (Gunell et al., 2019), and implicit particle in cell simulations have been seen to produce similar results (Deca

et al., 2019).

For highly active comets and on large scales, where an MHD description is adequate, the difference in electron and ion motion may be described by a Hall electric field (Huang et al., 2018).

### 1.2.3 Magnetic field carriers

The heavy ions in the cometary plasma are not magnetised, whereas the electrons are. Additionally, it has been shown that electron-scale physics are important even on larger scales (Deca et al., 2017).

The solar wind ion cavity is purely a region devoid of solar wind ions, not of solar wind electrons and not of the solar wind magnetic field. Thus, understanding **the flow of the electron fluid** is instrumental in understanding the behaviour of the magnetic field.

**The particle signatures at the diamagnetic cavity** are of particular interest. Cold electrons are present in the diamagnetic cavity and to a lesser degree outside of it (Odelstad et al., 2018). It has been observed that a suprathermal electron population associated with the solar wind is present just outside the cavity, but not inside (Madanian et al., 2017; Nemeth et al., 2016). Consequently, as the solar wind flow is mass-loaded the protons decouple from the field first, forming the solar wind ion cavity. Then at the diamagnetic cavity boundary, the magnetic field is stopped and so are the associated electrons.

### 1.2.4 Waves

Plasma waves take on an important role in the cometary plasma environment, **transferring energy** across boundaries and **heating particle populations** through wave–particle interactions. Waves are also instrumental in **setting up plasma boundaries** around the comet, e.g. the bow shock is formed when the relative velocity between the solar wind and the cometary plasma exceeds the wave speed (Coates, 1995).

A wide variety of plasma waves were detected starting millions of kilometres from the nucleus down to the closest approach at approximately 8000 km for the ICE spacecraft at 21P and the VEGA-2 spacecraft at Halley (Scarf, 1989; Tsurutani, 1991). **Ion acoustic waves** were detected by the ICE spacecraft during its traversal of the bow shock region at 21P (Scarf et al., 1986), and by the Sakigake spacecraft in the foreshock region upstream of Halley's comet (Oya et al., 1986). The Rosetta spacecraft observed ion acoustic waves both before the formation of the diamagnetic cavity (Gunell et al., 2017b) and later, when the cavity had formed, such waves were seen to be confined inside the cavity (Gunell et al., 2017a). In the plasma outside of the diamagnetic cavity, wave activity in the **lower hybrid frequency range** is abundant (André et al., 2017; Karlsson et al., 2017), and waves in this frequency range have also been observed at the boundary of the diamagnetic cavity (Madsen et al., 2018), indicating that a **mode conversion** from lower hybrid to ion acoustic waves takes place at the boundary. One of the first discoveries by the Rosetta spacecraft was the **"singing comet" waves** that were found at low frequencies (about 40 mHz) (Richter et al., 2015, 2016). These waves have been shown to be compressional (Breuillard et al., 2019), and they have been interpreted as the result of a modified ion-Weibel instability (Meier et al., 2016).

**Mirror mode** structures were observed in the magnetosheath of comet 21P (Tsurutani et al., 1999) and on both sides of the magnetic pileup boundary of comet 1P/Halley (Glassmeier et al., 1993). **Ion cyclotron waves** at the gyro frequency of water group ions were observed at comets 21P (Smith et al., 1986), 1P/Halley (Glassmeier et al., 1989; Yumoto et al., 1986) and 26P (Glassmeier and Neubauer, 1993; Neubauer et al., 1993). While both ion cyclotron and mirror mode waves were prominent during these comet encounters, they have, so far, eluded detection by the Rosetta spacecraft at its comet.

Thus, a wide variety of waves were observed in the fast flybys in the 1980s and 90s; recently the Rosetta mission has continued to find new plasma wave modes, and the waves have been seen to be linked to boundaries such as the bow shock and the diamagnetic cavity boundary. Still there are differences between the comets, and these are largely unexplained at this time.

### 1.2.5 Energetic Neutral Atoms

Energetic neutral atoms (ENA) are created when an energetic ion undergoes a **charge-exchange reaction** with a neutral atom or molecule, creating an energetic neutral atom (or molecule). **Charge exchange processes may remove much of the charge of the solar wind at a comet,** producing a neutral solar wind that may strike the inner collisional coma or the nucleus (Nilsson et al., 2015; Simon Wedlund et al., 2016). Charge exchange reactions with the solar wind are sometimes a significant source of ionisation of the coma (Simon Wedlund et al., 2017).

Charge exchange between cometary ions and neutrals is the most important collisional process in the marginally collisional coma, acting to slow down the ions while creating a component of low energy ENAs (Vigren and Eriksson, 2017).

No ENA instrument has been flown to a comet yet. Ekenbäck et al. (2008) conducted MHD simulations of a comet and found that remote observations at large distances should be feasible. **A comet shines bright in ENAs.** Charge exchange products can also be seen by ion instruments in the form of He$^+$ ions produced from solar wind He$^{2+}$ (Simon Wedlund et al., 2016; Simon Wedlund et al., 2019a,b). These observations can be used to independently assess the integrated **column density of the neutral atmosphere upstream** of the observation point (Hansen et al., 2016; Simon Wedlund et al., 2016).

Other missions have measured ENA emissions from objects that are similar to comets. For example, the Interstellar Boundary Explorer (IBEX) observed ENAs produced at the outer boundary of the heliosphere, the heliopause region (McComas et al., 2009), which has many similarities to the situation at a comet on a much grander scale. Two plasma streams meet (the solar wind and the ionised part of the interstellar medium) in a region with a significant neutral gas background. IBEX has also been used to obtain an overall image of the plasma structures of the Earth's magnetosphere (Fuselier et al., 2010). Another example relevant to comet studies are the observations of the subsolar magnetopause ENA jet at Mars. This jet is affected by the solar wind pressure, and that raises the possibility of a continuous remote monitoring of the effect of the solar wind on a magnetosphere (Futaana et al., 2006a). The ENA jet at Mars showed a periodic oscillation after the impact of an interplanetary shock passage, indicating that an oscillation of the boundary was excited (Futaana et al., 2006b).

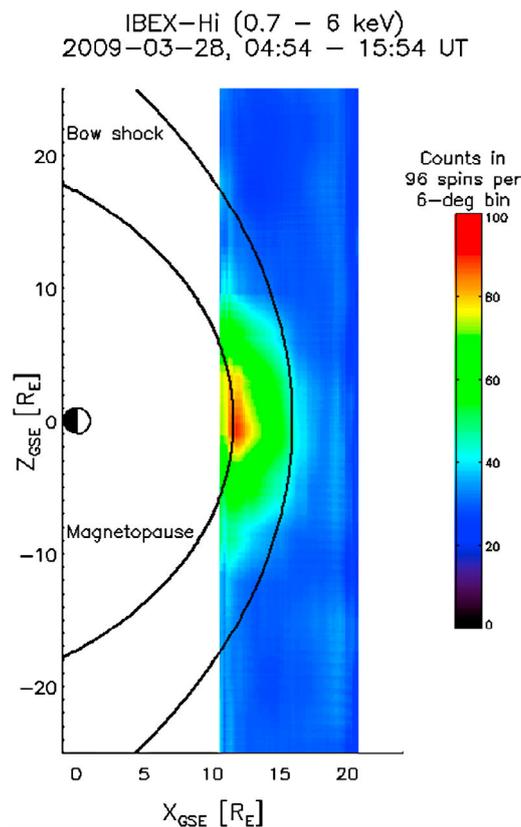

Figure 1.6: *ENA emissions from the subsolar magnetopause of Earth (Fuselier et al., 2010).*

### 1.2.6 Influence on the nucleus

The solar wind can directly influence the nucleus by solar wind ions hitting the surface, if the comet atmosphere is not too dense. **Solar wind sputtering of the surface** can release elements like C, O, Na, K, Si, Ca and S, which are less volatile than the typically released compounds $H_2O$, CO and $CO_2$. These less volatile materials were detected by ROSINA on Rosetta (Wurz et al., 2015), and were seen released from different areas of the nucleus than the volatile species. This could be due to the deflected solar wind hitting different parts of the nucleus than the sunshine or the lower degrees of attenuation of solar wind protons above the hemisphere of lower activity where the sputtered species have been observed. The release of surface materials by sputtering can be calculated through models (Ziegler, 2004), and thus the chemical composition of major elements of the areas affected by sputtering inferred from gas composition measurements (Schaible et al., 2017).

Sputtering often releases metals, which are ionised quickly and recombine slowly. They may therefore form long-lived metal ions, like the sporadic E layers observed in the ionosphere of Earth (Kirkwood and Nilsson, 2000).

Energetic molecules hitting the surface may also participate in **surface reactions**, thus affecting the chemistry at the nucleus. Yao and Giapis (2017) for example suggested that Eley-Rideal reactions could be the source of $O_2$ detected at comet 67P. Heritier et al. (2018a) later showed that this was not a plausible explanation at comet 67P, but it could still be a relevant mechanism for formation of other molecules. Sputtering is in general important for icy surfaces, not only at comets but also at icy moons (Johnson, 1998).

## 1.3 A review of dust–plasma interactions

None of the instruments on previous comet missions were well suited to investigate how the plasma and dust interact, so many open questions remain. The cometary environment is a region where physics of dusty plasmas is important and accessible to in situ study by visiting spacecraft (Mendis and Horányi, 2013; Vigren et al., 2015). Studying the dusty plasma at comets is also relevant to other bodies, e. g. at Enceladus (Boice and Goldstein, 2010). There, Cassini observations indicate that only a small fraction of the electrons escape attachment to dust grains and the dust consequently is of major importance for the plasma dynamics (Engelhardt et al., 2015; Morooka et al., 2011).

However, the **dust size distributions** in the two environments differ significantly, and so does the relative importance of dust-plasma interactions. Describing this distribution as a power law, the spectral index is approximately $4 - 5$ at Enceladus and in the E-ring (Kempf et al., 2008; Kurth et al., 2006). At 67P, Rosetta found a less steep dust distribution, with a spectral index ~ 3 for grain masses below 1 mg (corresponding to mm size), increasing to 3.6 post-perihelion (Fulle et al., 2016). Thus, for the same dust mass in a unit volume of space, less electrons attach to the dust grains as many small grains have much higher total capacitance than one large one of the same mass and the voltage they can charge to is limited by the kinetic energy of electrons in the plasma and therefore the dust-plasma interaction is weaker (Engelhardt et al., 2015). There are few Rosetta observations of dust grains below $\mu$m size, and subunits of larger grains have been found down to about 0.1 $\mu$m (Mannel et al., 2019).

Large dust grains break up into smaller fragments due to electric forces, stemming from the electric charge of the grain (Hilchenbach et al., 2017). Thus, not only does the size distribution influence the grain charge as described above, but **there is also an influence of the grain charge on the size distribution through fragmentation of dust grains**. As a result, if the plasma is well characterised and the charging processes are known, the dust size distribution will provide **information about the cohesive strength of the grains**.

Charged dust grains in the sub-micrometre size range are moved by electromagnetic forces and, in regions where the gas density is high, also by the drag force from the neutral gas. Charged nano-grains were detected by the electron spectrometer onboard the Rosetta spacecraft (Burch et al., 2015). Due to the small charge-to-mass ratio, charged nano-dust trajectories have a large radius of curvature, and they are approximately parallel to the electric field. Rosetta results have shown that the electric field around a comet is highly structured (Section 1.2.2), which affects the motion of the charged dust. For example, the ambipolar electric field would act to confine neg-atively charged grains to the inner coma. The interaction also goes the other way: the collection of electrons on the much heavier dust grains affects how the electric field is structured in the coma, and the presence of dust influences the wave modes in the plasma, such as dust acoustic and dust ion acoustic waves (Barkan et al., 1995; Merlino, 1997). Figure 1.7 shows dust acoustic waves in the laboratory as an example of dust–plasma interactions.

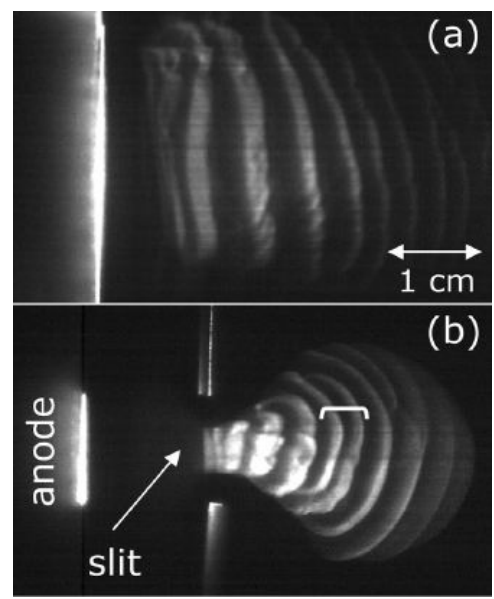

Figure 1.7: *Dust acoustic waves (Heinrich et al., 2009).*

# 2 Future science questions

## 2.1 Large scale structures in the interaction region

### 2.1.1 Bow Shock

Simulations have shown that the location of the bow shock is asymmetric along the direction of the solar wind convective electric field (Koenders et al., 2013) to a greater extent than what has been seen at Mars (Mazelle et al., 2004), and this asymmetry is highly dependent on cometary activity and solar EUV intensity (Lindkvist et al., 2018). Not all flybys of comets show a well defined shock, sometimes only a bow wave was observed (Neubauer et al., 1993). It was not possible to observe a fully developed bow shock with Rosetta, due to spacecraft trajectory constraints. This also means that the shape of a cometary bow shock has never been observed. At planets, data from different locations, probed during many spacecraft orbits, has been used to form a statistical picture of a mean bow shock shape. While a snapshot of the bow shock shape at one moment in time would require multi-point measurements, single-point observations can be used to determine what the bow shock shape is on average if the data set covers a sufficiently large range of positions. At comets, all we have so far are single point measurements in flybys and Rosetta observations from a small region over which an infant bow shock moved at different times (Gunell et al., 2018a). Therefore, the bow shock shape, width and structure have not been adequately determined and are largely unknown.

In modelling, the shape of a bow shock and its width are direct consequences of the assumptions on which the models are based. Verifying the shape observationally is therefore important for our understanding of the physics governing the formation of bow shocks at comets. This is particularly true in the case of a bow shock under formation, that is to say, an infant bow shock, which cannot be studied at solar system objects of any other kind. Therefore the main question is

⇒ **How asymmetric are large-scale structures at comets?**

The infant bow shock as it was observed was not always constant, instead changes could be seen on varying timescales (Gunell et al., 2018a). What is driving these changes? Are they driven by changes in the upstream solar wind, by variable outgassing from the nucleus or is the bow shock itself unstable, leading to waves that are seen as variable conditions by a stationary observer? At Earth bow shock ripples have been observed in multi-spacecraft studies (Johlander et al., 2018). Bow shock ripples are thought to be a cause of high speed jets (or plasmoids) in Earth's magnetosheath, and such jets have been found to be geoeffective (Plaschke et al., 2018). Do cometary bow shocks support surface waves and ripples? Can ripples lead to jet formation at comets, and if so what would the impact of those jets be on the plasma and coma downstream of the shock? How do these bow shock properties change as the bow shock transforms from an infant to a fully developed shock? The answers to these questions will have an impact on our understanding not only of cometary bow shocks, but also of both the comet–solar wind system as a whole and of the physics of bow shocks in general.

What heats the plasma as it passes the shock? Is it heated by reflection followed by thermalisation or by waves excited by plasma instabilities? Both these scenarios are known to occur at Earth (Bale et al., 2005; Eastwood et al., 2005). At a comet the situation is more complicated than at a planet, since in the vicinity of a cometary bow shock there are both cometary ions and neutrals present. Additional phenomena seen at Earth's bow shock that remain unexplored at comets include electric fields at the shock that may contribute to charged particle reflection, acceleration, and heating; field-aligned particle beams; and foreshock waves in a variety of frequency ranges. These questions form part of the more general problem of how mass, energy and momentum are transferred in the cometary environment, through the coma and across boundaries.

### 2.1.2 Solar Wind Ion Cavity

It is known from both the Rosetta observations and previous in situ measurements, in fast flybys of comets, that once a comet is active enough a boundary which demarcates the region that solar wind ions cannot penetrate is formed. Other boundaries have been identified in this region of the plasma that could not be found in the Rosetta observations. Therefore, we ask:

⇒ **Which boundaries exist at a comet during its journey through the solar system?**

Is the solar wind ion cavity the only boundary in the cometosheath? What causes the formation of this boundary? How important are different physical processes, such as mass-loading, magnetic pileup, ionisation processes, and the wide variety of collisional processes at work at a comet? These questions remain unanswered today, and answering them would require multi-point measurements to determine how the boundary is structured as well as quantitative observations of collisional processes in the coma. Additional information can be provided by ENA measurements, observing the main regions of a comet magnetosphere remotely. The relative importance of the mechanisms involved is likely to change during the course of a comet's orbit around the Sun. Thus, conducting multi-point measurements at a range of heliocentric distances, we can advance our ability to predict boundary properties under varying conditions. Such knowledge will be of significance to planetary studies, including exoplanets, since at unmagnetised planets a corresponding boundary, the induced magnetosphere boundary, is responsible for protection from atmospheric escape (Gunell et al., 2018b).

### 2.1.3 Diamagnetic Cavity

The main open question that remains with regards to the diamagnetic cavity is

⇒ **What are the processes behind the diamagnetic cavity formation? Is it different for 67P and 1P?**

The mechanism of cavity formation is still poorly understood, with theories diverging for the two comets at which this region has been observed.

Unfortunately, Rosetta was not able to measure the distribution function of the lower energy ions (Masunaga et al., 2019) due to a very negative spacecraft potential. These ions are instrumental in understanding the plasma dynamics at the boundary as their direction and speed can give insight into the particle dynamics at the boundary and their interaction with the neutral gas. This in turn will provide more information on the diamagnetic cavity formation mechanism.

The distribution function of the electrons is also poorly understood. The interplay of newly created warm photo-electrons, cold electrons and suprathermal electrons has not been investigated in detail and available data is severly lacking in accuracy and temporal resolution. New observations with higher temporal resolution, better angular coverage and at low spacecraft potential are needed to understand these dynamics.

Furthermore, the true shape of the boundary has not been measured, as this requires measurements at at least two points of the boundary at the same time. It remains to be investigated with the help of multi-point measurements what the exact nature of the boundary oscillations is. It is unclear how these oscillations are affected by a change in the gas production rate. Are the oscillations large and fast enough to explain the quick succession of diamagnetic cavity encounters at 67P?

The situation at both Mars and Venus should be very similar to the one at the comet, but no observations of a completely field free region have been reported there. However, it should be noted that the distance at which this boundary might be detected is most of the time below the spacecraft trajectories.

### 2.1.4 Plasma Tail

Tail disconnections have been revealed by remote optical observations, and tail rays can also be seen in pictures of comets. In the absence of comprehensive in situ measurements from comet tails, all that exist are a few fast fly-throughs, the mechanisms behind these phenomena are as yet unknown.

⇒ **What is the cause of tail disconnection events and tail rays?**

Various theories have been proposed to explain tail disconnection events (see Section 1.1.4), but none of these give a complete and satisfactory answer. Answering this question will, in turn, teach us a great deal about the cometary plasma environment. Is reconnection a relevant concept in comet plasma physics? What causes ionisation and plasma acceleration in comet rays? And why are there rays at all, as opposed to a uniform expansion in all directions? Furthermore, observing the comet tail plasma in situ will reveal how the tail is structured, what plasma instabilities are present, and how this compares to the plasma tails of unmagnetised planets.

## 2.2 Plasma processes at a comet

### 2.2.1 Collisions in the coma

A comet presents an excellent opportunity to monitor the collisionality and evolution of a partially ionised environment. Due to the elliptical orbits of comets, the ion coma evolves and transitions between the collisional and collisionless regimes. This will help us understand how collisions compete with other processes on the microscopic level and how these effects influence a large scale system, that is to say, the comet as a whole.

⇒ **What is the role of collisions in the densest part of the coma?**

Despite past and recent sustained experimental efforts, many relevant cross sections for charge-changing and ionisation collisions, involving water or other abundant species such as CO and $CO_2$ at energies below 1 keV, are not known with an accuracy sufficient to support accurate modelling of solar wind–cometary interaction. Also, the precise energy distribution of cross sections can play an important role when convolved with a heated solar wind ion distribution. Therefore, new extended laboratory experiments are needed to better constrain these cross sections and their shape at relevant energies. In turn, the investigation of the plasma composition at a comet can help constrain cross sections that are not accessible in the laboratory.

At low outgassing activity conditions ($Q < 5 \times 10^{26}\,\mathrm{s^{-1}}$) cold electrons were observed at 67P (Engelhardt et al., 2018; Gilet et al., 2017). However, radial energy degradation models cannot explain the significant cooling of the newly-born electrons. The complex electro-magnetic environment, as suggested by large scale simulations (e.g., Deca et al., 2017), may contribute to the energy budget of the cometary electrons.

At high outgassing activity conditions ($Q > 5 \times 10^{27}\,\mathrm{s^{-1}}$), near perihelion at comet 67P, the diamagnetic cavity was observed near the electron exobase (Henri et al., 2017), where the electrons transition between the collisional and collisionless regimes. The formation mechanism of the diamagnetic cavity is a question in itself, and it is not known what role collisions may have in it.

For intermediate and high outgassing activity conditions ($Q > 10^{27}\,\mathrm{s^{-1}}$), the ion composition changes and becomes richer as comets get closer to the Sun, which is a piece of evidence of ion-neutral collisions taking place (Haider and Bhardwaj, 2005; Heritier et al., 2017a). Rosetta observed hourly and daily variations of the triplet $H_2O^+/H_3O^+/NH_4^+$ at comet 67P, ruling out the idea of a steady-state ionosphere (Beth et al., 2016), but the reason behind the variability remains an open question.

A magnetised plasma streaming through a neutral background — like the solar wind streams through the coma — is the setting for the modified two-stream instability behind the "Critical Ionisation Velocity" hypothesis suggested by Alfvén (1954). While kinetic energy is transferred from ions to electrons via a plasma instability, the actual ionisation happens in collisions between the energised electrons and the neutrals. The phenomenon has been observed in laboratory plasmas, but it has so far eluded detection in space (Lai, 2001).

We may formulate a number of specific science questions of importance on this topic (non-exhaustive list): (i) what is the role of charge-changing reactions in the local and large-scale dynamics (time scales, ion trajectories) of the plasma? (ii) how is energy transferred from the solar wind to the cometary plasma and neutral environment? (iii) how are the plasma boundaries formed at comets and what precise role do collisions play? (iv) how are electrons cooled in the inner coma? (v) why is the ion composition so variable? (vi) how stable is the ion–neutral two-stream interaction in a coma environment? and (vii) are negative ions abundant and what is their role? All these questions highlight the complicated interplay of collisional and electromagnetic processes in the cometary plasma.

Systematic investigation of charge-exchange effects with ion and ENA instruments concomitantly probing the cometary plasma, concentrating on the 3-D distribution of these species, will help shed light on these aspects. Moreover, the scope of all such studies far exceed the sole dominion of comet-solar wind interactions – our understanding of planet-solar wind environments as well as that of other astrophysics environments (interstellar medium, etc.) will benefit from them.

### 2.2.2 Electric fields

The three major contributions to the DC electric fields at comet 67P have been identified following the Rosetta mission. At that comet, the DC fields were relatively more important than ion–neutral collisional coupling, in contrast to the results obtained at comet 1P. Thus, to understand the behaviour of the plasma, which is affected by the fields, we need to understand how the plasma interacts with the neutrals. In particular, detailed measurements of charged particle distributions around the comet will be necessary to enable us to understand how the fields are generated. The charged particle distributions are the source of the electric fields, and the fields affect the particle distributions. Thus, the formation of the electric fields is intimately linked to the effect of these same fields.

⇒ **How do electric fields contribute to energy, momentum and mass transfer in the plasma?**

The formation of electric fields in the coma, as a result of interaction with the solar wind, affects the dynamics of both the cometary plasma and charged dust, transferring mass in the coma and tail. This, in turn, affects both tail properties and extended sources of gas released from the dust particles. Multi-point measurements of plasma, fields and dust will elucidate the physics behind mass transfer and the consequences for both the coma and tail.

This topic is not limited to the large scale DC electric fields. Fields on small scales at the plasma boundaries, such as the bow shock and the diamagnetic cavity boundary, are likely to play a significant part in forming and maintaining these boundaries. Measuring these fields will advance our understanding of boundaries at comets.

### 2.2.3 Frozen-in condition and magnetic field carriers

In the solar wind, both electrons and ions are magnetised. In the diamagnetic cavity of a comet, neither ions nor electrons are. In between there is a region where the electrons are magnetised but the ions are not. The behaviour of the electrons at the interfaces between these regions is as yet unknown.

⇒ **What is the role of the electrons as magnetic field carriers in a plasma where the ions are not magnetised?**

The magnetic field is frozen-in to the electrons rather than the ions. There are regions with electrons of both solar wind and cometary origin and with several electron populations of different temperatures. If the various electron populations behave similarly with respect to the magnetic field, it is yet unclear to which electron population the magnetic field is frozen-in.

The problem of magnetisation is closely related to other questions about the plasma boundaries. What keeps the electrons outside the diamagnetic cavity from entering it? How is the current that maintains the difference in magnetic field intensity across the diamagnetic cavity boundary generated and maintained? What prevents the solar wind ions from entering the solar wind ion cavity, while allowing the solar wind electrons to pass through?

### 2.2.4 Waves

Previous space missions have found that waves are ubiquitous in the comet environment. However, it is not clear in which region of the coma these waves are present and how they depend on activity level of the comet. Through multipoint measurements in the coma one can determine the temporal and spatial development of the waves. The details of wave propagation and the role of waves in diamagnetic boundary physics are not well understood. Going from single-spacecraft to multi-spacecraft observations will enable new insights into both the physics of the waves themselves and how they affect boundaries and the surrounding plasma. It has been speculated that wave–particle interaction and particle collisions transfer energy from the solar wind to the cometary plasma and redistribute energy in the coma.

⇒ **How do waves and wave–particle interactions affect the cometary plasma?**

Wave measurements are a necessary part of the assessment of how energy is transferred in the cometary environment.

Waves on electron timescales, that is to say, near frequencies typical for electrons, such as the plasma frequency, are important as they influence the electron distribution function, and dissipate energy in the cometary plasma. At comet 67P, hot electron populations were observed outside the diamagnetic cavity, and at the infant bow shock. Capabilities to sample waves at the plasma frequency will enable measurements of this family of waves, and to further the understanding of energy conversion on electron scales.

Waves on ion time scales have been associated with the bow shock at comets 1P and 21P and with the diamagnetic cavity boundary at comet 67P. It has been proposed that the waves are driven by currents that flow at these boundaries. Therefore, wave measurements can aid the understanding of the boundaries themselves, and shine a light on both how plasma particles generate waves in the cometary plasma and how these waves contribute to heating of the particle populations.

The low frequency waves that have been observed: singing comet, mirror mode, and ion cyclotron waves are in principle understood in terms of plasma theory. However, the differences between the Rosetta observations at comet 67P and what was observed during the flybys of comets 1P, 21P, and 26P have not yet been completely explained. For example, why have no ion cyclotron waves been detected by Rosetta? The role of low frequency waves at comets, and how they are generated depending on cometary properties is an open question.

In dust–plasma relations, we know from observations that there is a distributed source of certain species (De Keyser et al., 2017; Dhooghe et al., 2017) and hence that there is a significant amount of dust in the coma. Charges bound to heavy dust particles give rise to new wave modes in the plasma (Barkan et al., 1995; Merlino, 1997) and the detection of these waves provides an alternative measurement of the dust content.

### 2.2.5 Influence on the nucleus

That the nucleus affects the plasma in the coma is obvious, since outgassing from the nucleus is the source of the coma. However, the plasma can affect the surface of the nucleus through sputtering and chemical reactions, thus changing the composition of the emitted gases. The plasma also has an influence on the charging of dust and of the surface of the nucleus itself.

⇒ **How do the plasma and the nucleus interact?**

How do particle fluxes to the nucleus affect the composition of the emitted gases? Whether detected gases have been embedded in the nucleus for billions of years or formed recently in surface–plasma interactions is a piece of information necessary in interpreting observations of these gases. This will help us to assess where, when and how compounds found in the coma, such as $O_2$ (Bieler et al., 2015), formed — on the surface of cometary nucleus, on dust grains in the protosolar nebula, or elsewhere?

In order to be able to assess what processes are active on the surface we need to know the fluxes of energetic neutrals and ions onto it. This will enable modelling of the effect of sputtering on the surface, and it is also needed to infer surface composition from observations of sputtered or otherwise released or created material. Measuring the flux of electrons to and from the surface will enable us to determine the surface charge, and this quantity also affects dust levitation through charging.

## 2.3 Dust-plasma interactions

Dust in the coma interacts with the plasma charging the dust grains either positively or negatively. Dust motion in the coma is affected by both dust grain and plasma properties, and dust grains may constitute a distributed source in of gas emissions in the coma as was seen at comet 67P for the halogens (De Keyser et al., 2017). While we know what forces can act on a dust grain, and the basic processes for charging of the grains are known, the behaviour is expected to be very different for different grain sizes and different plasma parameters. Therefore, we still need to ask a basic question.

⇒ **What is the role of charged dust in the coma?**

How is the dust distributed in space around a comet? The spatial distribution is affected by the electric fields, and if the dust content is sufficiently high, charged dust will have a significant effect on the electric field.

What are the size and charge distributions? As charging of grains can lead to fragmentation, the space and charge distributions provide information of the grains themselves.

How does the presence of dust affect the sources of neutral gas and plasma in the coma? If the dust density is high, outgassing from the grains may be a significant source of neutral gas. In that case, the motion of dust grains under influence of electromagnetic forces may have an appreciable effect on neutral gas observations.

How do plasma waves interact with the dust? Observations of dust waves in the plasma can provide an indirect means of assessing the dust content in the coma.

Observations of nano-dust may be performed indirectly via electron, ion and wave measurements; large grains may be observed optically; and intermediate grain sizes require dedicated dust detectors. Since spacecraft more often than not are charged to a potential different from the ambient plasma, charged dust grains must overcome the potential barrier to reach the spacecraft and be detected. When the relative speed between the spacecraft and the comet is low, as in the Rosetta case, this represents a challenge in measuring the low energy dust, and it may require development of new experimental techniques. Possible directions for the development to take include putting a dust detector on a long boom on a spinning spacecraft, thereby increasing the detector to dust relative speed, or controlling the potential of the dust detector to enable the charged dust to reach it. The ability to correct for spacecraft potential fluctuations is valuable not only in dust detection, but also for accurate measurements of low energy electrons and ions.

# 3  Possible missions

In order to answer the science questions presented in section 2.1, two aspects of space missions are particularly important: **conducting multi-point measurements** and **orbiting a comet for an extended period of time**. This is necessary to obtain the 3D structure, differentiate temporal and spatial variability, and simultaneously assess the variation in different plasma regions. Ideally, a mission should be able to do all of that to achieve the highest science return. We propose three different mission profiles that, at least partially, address the science questions. As can be seen from Table 4.1, only mission profile A is able to answer all science questions, whereas B and C focus on specific subsets.

## A: Multi-spacecraft mission

This mission follows a comet in the same way Rosetta did: it will rendezvous with a comet well before it reaches perihelion, and accompany it for as long as possible. The new concept compared to Rosetta is that it is optimized for plasma measurements and consists of several identical spacecraft and a mother spacecraft. This ensures that simultaneous, inter-comparable measurements at multiple points can be performed.

Although the number of spacecraft is theoretically unlimited, we suggest to have at least four in total, which enables us to use methods that have been tried and tested using missions such as Cluster, Themis and MMS. For example, the curlometer technique can give 3D measurements of the current in the region between the spacecraft, a measurement that is important for almost all the science questions.

The trajectory at a comet is easily adjustable, as the gravitation of the nucleus is very small and thus the formation of the spacecraft can be changed regularly. This means that it can be switched from a tetrahedron (for the curlometer technique to determine 3D currents) to a pearls-on-a-string formation with large cometocentric distance variations which is advantageous to measure the oscillations along a boundary.

To answer the main science questions, the following quantities need to be measured:

- 3D magnetic field
- 3D electric field
- plasma density, temperature and heat flux
- ion velocity distribution function with mass resolution (at least protons/water/carbondioxide)
- ENAs
- electron velocity distribution function
- neutral gas density and major constituents
- dust flux
- visible light and UV images of nucleus and coma

Thus, the proposed instrument suite would include a scalar and vector magnetometer and an electric field instrument to measure the fields. An ion mass/energy spectrometer with an ENA detector and an electron energy spectrometer should then provide detailed moments of the particle distribution functions. To support the plasma measurements and in particular the monitoring of the interactions between the ionised and the neutral phases that form a cometary coma, a neutral gas pressure gauge, a visible light camera and a UV camera should also be included. To support the monitoring of the interactions between the ionised and the dust phases, a nanodust detector should be included. All instruments have heritage from missions such as Rosetta, Venus Express, Mars Express, Cluster, Cassini, Bepi-Colombo and JUICE. However, new development is necessary to achieve the accuracies needed in some cases. For example, to assess the role of dust, the difference in ion and electron density must be determined even at low spacecraft speeds.

For calibration purposes and to cover the entire sky with the field of view, the spacecraft should be spinning. To enable measurements of the low energy ions, the spacecraft potential should be kept as close to zero as possible. This is challenging in a plasma that is warm and dense, as is the case at a comet, some further technological development in active spacecraft potential control is required. To optimise science return only the mother spacecraft could carry the remote observing instruments and high gain antenna, which means it should be three-axis stabilised. The spinning spacecraft would transmit their collected data to the three-axis stabilised one, which would handle communications with Earth. The radio-links between the spacecraft could also be used to measure the total electron content along a line between two spacecraft. Four spacecraft can be manoeuvred so that these lines of sight intersect, allowing tomographic analysis of the electron density.

## B: Sub-spacecrafts within a cometary mission

For this mission profile we assume that a mission to explore a cometary nucleus is planned in the near future. We propose one or multiple sub-spacecraft (spacecraft II) to accompany a Rosetta type comet mission (orbiter, spacecraft I). Spacecraft I should provide communication with Earth as well as the main propulsion system for the cruise phase. One of the sub-spacecraft could be provided by an international partner as was done for e. g. Bepi-Colombo.

To enable multi-point measurements the orbiter should also be equipped with a suite of plasma instruments similar to that on Rosetta, i.e. a magnetometer, a Langmuir probe and electron and ion sensors.

We propose one or multiple sub-spacecraft, dedicated to investigations of the plasma. This has the advantage that this spacecraft can also go to large cometocentric distances without impacting the main spacecraft's science goals. The instrument suite should be similar to that proposed in mission profile A. It is also assumed that spacecraft I is equipped with a neutral gas monitor and an imager for its main science goals. The assumption is that this spacecraft is as close as possible to the nucleus to enable detailed investigations of the surface.

## C: Artificial comet

To investigate the fundamental properties of plasma physics, space could be used as a convenient, accessible, clean physics laboratory. For instance, the impact of heavy ions on a plasma flow can be directly investigated, in a controlled way, by creating an artificial comet near Earth. This has the advantage of high telemetry rates and low propulsion requirements. The spacecraft should be equipped with similar instruments as mission profile A with the addition of an assembly to deploy canisters of heavy atoms/molecules. The canisters are then exploded and the ionisation and subsequent incorporation of the heavy ions into the solar wind could be studied in more detail. In order to obtain multi-point measurements the release should be coordinated with other Earth-orbiting satellites so that these also pass through the cloud. There are currently several plasma-instrument carrying missions available, e. g. Cluster, Themis and MMS. The event could also be observed remotely with either ground- or space-based telescopes.

Such as experiment would be very similar to the AMPTE mission, with the added advantage of existing infrastructure in the near-Earth solar wind and much improved instruments on the main spacecraft. There is also a possibility of observing the artificial comet with telescopes from Earth with involvement of the amateur astronomy community. The choice of gas should also be adjusted to better reflect the ionisation rate at real comets, because Barium (used by AMPTE) has an ionisation frequency that is three orders of magnitude larger than that of water.

## Additional measurements

To support the three mission profiles other measurements will be beneficial. For example, laboratory measurements of charge exchange rates or collision rates are still needed to provide the input parameters for detailed models of the plasma. Energy-dependent collisional cross sections for low energy particles are of particular interest. Also, experiments on ice sublimating in a vacuum will aid the understanding of the development of dust grains containing a combination of ice and refractory materials.

   Ground based remote observations of plasma tails are needed to provide large scale context for the in-situ measurements. For ideal coverage the amateur astronomy community should be included. New telescopes, both Earth and space based, will increase the detection rate for comets and provide an even larger catalogue of mission targets.

   Simulations, multi-fluid MHD, hybrid, or fully kinetic on large scales are needed to provide the necessary context to interpret the in-situ observations.

# 4 Conclusions

Table 4.1 summarises the main science questions outlined in this white paper. The investigation of the cometary plasma environment will not only provide new insights on the interaction of a comet with the solar wind but can also be a useful vehicle to study the impact of small scale plasma processes on large scale structures. Of the three mission profiles shown here, the multi-point rendezvous mission (A) is needed to answer many of the questions that still remain.

   Especially after the unprecedented and unequalled success of the European missions to comets, Giotto and Rosetta, as well as the upcoming Comet Interceptor, we urge ESA to keep up their predominance in this sector of space research and include a cometary plasma mission in the Voyage 2050 program.

| Science Question | Mission profile |
|---|---|
| How asymmetric are large-scale structures at comets? | A, B[†] |
| Which boundaries exist at a comet during its journey through the solar system? | A, B |
| What are the processes behind the diamagnetic cavity formation? Is it different for 67P and 1P? | A, B*, C |
| What is the cause of tail disconnection events and tail rays? | A |
| What is the role of collisions in the densest part of the coma? | A, B, C |
| How do electric fields contribute to energy, momentum and mass transfer in the plasma? | A |
| What is the role of the electrons as magnetic field carriers in a plasma where the ions are not magnetised? | A, C |
| How do waves and wave–particle interactions affect the cometary plasma? | A |
| How do the plasma and the nucleus interact? | A, B |
| What is the role of charged dust in the coma? | A, B |

*Spacecraft I and II need to be in the boundary region at the same time.
†Spacecraft I has to provide radial coverage.

Table 4.1: Main science questions and mission profiles suggested to solve them.

# 5 Supporters

**Charlotte Götz\***
Technische Universität Braunschweig, Germany

**Herbert Gunell\***
Royal Belgian Institute for Space Aeronomy, Belgium and Umeå University, Sweden

**Martin Volwerk\***
Institut für Weltraumforschung, Austria

**Arnaud Beth\***
Imperial College London, United Kingdom

**Anders Eriksson\***
Swedish Institute of Space Physics, Uppsala, Sweden

**Marina Galand\***
Imperial College London, United Kingdom

**Pierre Henri\***
LPC2E, CNRS, Orléans France and Laboratoire Lagrange, OCA, CNRS, UCA, Nice, France

**Hans Nilsson\***
Swedish Institute of Space Physics, Kiruna, Sweden

**Cyril Simon Wedlund\***
Institut für Weltraumforschung, Austria

**Markku Alho**
Aalto University, Finland

**Laila Andersson**
LASP, University of Colorado, United States

**Nicolas Andre**
IRAP, Toulouse, France

**Johan De Keyser**
Royal Belgian Institute for Space Aeronomy, Belgium

**Jan Deca**
LASP, University of Colorado, United States

**Yasong Ge**
Chinese Academy of Sciences, China

**Karl-Heinz Glaßmeier**
Technische Universität Braunschweig, Germany

**Rajkumar Hajra**
National Atmospheric Research Laboratory, India

**Tomas Karlsson**
Royal Institute of Technology, Stockholm, Sweden

**Satoshi Kasahara**
University of Tokyo, Japan

**Ivana Kolmasova**
Czech Academy of Sciences, Czech Republic

**Kristie LLera**
Southwest Research Institute, United States

**Hadi Madanian**
The University of Iowa, United States

**Ingrid Mann**
UiT, The Arctic University of Norway, Norway

**Christian Mazelle**
IRAP, Toulouse, France

**Elias Odelstad**
Royal Institute of Technology, Stockholm, Sweden

**Ferdinand Plaschke**
Institut für Weltraumforschung, Austria

**Martin Rubin**
Universität Bern, Switzerland

**Beatriz Sanchez-Cano**
University of Leicester, United Kingdom

**Colin Snodgrass**
University of Edinburgh, United Kingdom

**Erik Vigren**
Swedish Institute of Space Physics, Uppsala, Sweden

★: Contributor